\RequirePackage{fix-cm}
\documentclass[twocolumn,epjc3]{svjour3}  
\smartqed  
\RequirePackage{color,graphicx,url}
\usepackage{amsfonts,amsmath,amssymb,bm,xspace}

\newcommand{\tov}{\mathrm{TOV}}

\newcommand{\fmi}{\mathrm{fm}^{-1}}
\newcommand{\fmiq}{\mathrm{fm}^{-3}}

\newcommand{\ChiEFT}{\ensuremath{\mathrm{\chi EFT}}\xspace}

\newcommand{\mm}{\mathrm{MM}}

\newcommand{\sm}{\mathrm{SM}}
\newcommand{\nm}{\mathrm{NM}}
\newcommand{\ffg}{\mathrm{FFG}}

\newcommand{\tot}{\mathrm{tot}}
\newcommand{\std}{\mathrm{std}}
\newcommand{\ex}{\mathrm{exp}}
\newcommand{\refr}{\mathrm{ref}}

\newcommand{\nuc}{\mathrm{nuc}}
\newcommand{\cl}{\mathrm{cl}}

\newcommand{\sat}{\mathrm{sat}}
\newcommand{\sym}{\mathrm{sym}}

\newcommand{\bulk}{\mathrm{bulk}}
\newcommand{\fs}{\mathrm{FS}}
\newcommand{\coul}{\mathrm{Coul}}

\newcommand{\surf}{\mathrm{surf}}
\newcommand{\curv}{\mathrm{curv}}

\newcommand{\cc}{\mathrm{cc}}
\journalname{Eur. Phys. J. A}
\begin{document}


\title{Confronting a set of Skyrme and $\chi_{EFT}$ predictions for the crust of neutron stars
}
\subtitle{On the origin of uncertainties in model predictions}


\author{Guilherme Grams\thanksref{addr1}
        \and
        J\'er\^ome Margueron\thanksref{addr1} 
        \and
        Rahul Somasundaram\thanksref{addr1} 
        \and
        Sanjay Reddy\thanksref{addr2,addr3} 
}



\institute{Univ Lyon, Univ Claude Bernard Lyon 1, CNRS/IN2P3, IP2I Lyon, UMR 5822, F-69622, Villeurbanne, France \label{addr1}
           \and
           Institute for Nuclear Theory, University of Washington, Seattle, WA 98195-1550, USA \label{addr2}
           \and
           JINA-CEE, Michigan State University, East Lansing, MI, 48823, USA \label{addr3}
}

\date{Received: date / Accepted: date}

\maketitle

\begin{abstract}
With the improved accuracy of neutron star observational data, it is necessary to derive new equation of state where the crust and the core are consistently calculated within a unified approach. For this purpose we describe non-uniform matter in the crust of neutron stars employing a compressible liquid-drop model, where the bulk and the neutron fluid terms are given from the same model as the one describing uniform matter present in the core. We then generate a set of fifteen unified equations of state for cold catalyzed neutron stars built on realistic modelings of the nuclear interaction, which belongs to two main groups: the first one derives from the phenomenological Skyrme interaction and the second one from $\ChiEFT$ Hamiltonians. 
The confrontation of these model predictions allows us to investigate the model dependence for the crust properties, and in particular the effect of neutron matter at low density.
The new set of unified equations of state is available at the CompOSE repository.
\end{abstract}
%
%
\section{Introduction}
\label{intro}

The description of the neutron star (NS) equation of state (EoS) from the crust to the core represents a challenge for modern nuclear and particle physics, as well as for astrophysics~\cite{Lattimer:2004,LATTIMER2016127}, see also Ref. \cite{Burgio2021} for a review. A typical density at the transition between the crust and the core is about half saturation energy-density ($\rho_\sat\approx 2.6~10^{14}$ g~cm$^{-3}$) while in the core, it reaches up to several times this energy-density.
These densities determine NS global equilibrium properties, such as their masses, radii, moment of inertia or tidal deformabilities~\cite{Haensel:2007yy}. Recently, NICER X-ray observatory has released the measurement of two NS masses and radii with an unprecedented accuracy: PSR J0030 has been estimated to have a mass M$=1.4\pm0.05$~M$_\odot$ and a radius R$=13.02^{+1.24}_{-1.06}$~km~\cite{MillerNICER2019} or R$=12.71^{+1.14}_{-1.19}$~km~\cite{Riley19}, and PSR J0740 with a mass M$=2.07\pm0.05$~M$_\odot$ and a radius
R$=12.35\pm0.75$~km~\cite{Miller21} or R$=12.39^{+1.30}_{-0.98}$~km~\cite{NICER2021}.
Now measurable with LIGO-Virgo interferometers, the tidal deformability $\tilde{\Lambda}$ has been estimated for the first time to be $\tilde{\Lambda}=280^{+300}_{-200}$ from GW170817~\cite{Abbott2017}, the gravitational waveforms emitted during the last inspiral orbits of the binary NS mergers. The tidal deformability is in turn strongly correlated to the mass and radius~\cite{Tews:2018kmu}, converting the uncertainty on $\tilde{\Lambda}$ into an uncertainty on the radius of a 1.4M$_\odot$ NS of about $\pm 1$~km. The confrontation of theoretical modeling against accurate observational data, especially the very recent ones from NICER and the LIGO-Virgo-Kagra collaboration (LVKC), is now possible and represents a major challenge requiring more and more precise modeling of dense matter properties. 

The new precision era, opened by NICER and LVKC for the measurement of radii and tidal deformabilities, requires to control more systematically the sources of theoretical uncertainties in the modeling of NS EoS. One of them is the modeling of NS crust, which may be inconsistent with the one used for the core. While the crust represents a small fraction of the NS, 10\% in terms of radius and less than 1\% of the mass, it was estimated that the method used to connect the crust and the core could influence the theoretical prediction of the NS radius by 3-5\%, or in other words, by a few hundred of meters~\cite{Fortin2016}. See also Ref. \cite{Suleiman2021} for a recent study on the impact of non consistent treatment of crust and core EoS on NS macroscopic properties. Presently, the experimental uncertainties in the measurement of NS radii are still larger -- about 1-2~km -- than this theoretical one. However, anticipating future observational improvements, it is preferable to resolve this source of uncertainty by employing unified models from the crust to the core of NS, as it was already suggested by several teams~\cite{DouchinHaensel2001,Fantina12,Fortin2016}.


Even when NS EoS are unified from the crust to the core, they still lead to different predictions depending on the nuclear interaction on which they are built on. The scope of this paper is therefore to confront a set of unified EoS for cold catalyzed NS matter in order to evaluate the model dependence of the crust predictions. To do so, we employ a compressible liquid-drop model (CLDM) originally suggested by Baym, Bethe and Pethick~\cite{bbp1971}, and which we have recently applied to include state-of-the-art nuclear physics constraints, e.g. $\ChiEFT$ interaction used in Many-Body Perturbation Theory (MBPT) framework, as well as nuclear physics data such as the AME2016 database~\cite{AME2016}. In the CLDM, the EoS used in the core is unified with the crust, through the bulk and neutron fluid contributions, see for instance the recent Refs.~\cite{Carreau2019a,Grams:2021a} and references therein. The effects of non-uniformities in the crust are implemented through finite-size (FS) terms. In Ref.~\cite{Grams:2021b}, we have shown that these terms can be sorted according to the leptodermous expansion~\cite{Myers1973}. We have also presented in detail the use of the meta-model to capture the predictions of the $\ChiEFT$ Hamiltonians. A first comparison of the $\ChiEFT$ predictions against a representative phenomenological model, SLy4~\cite{Chabanat1997}, was performed in Ref.~\cite{Grams:2021a}. In this paper, we perform a more extensive comparison considering a set of seven Skyrme parametrisations which are used to determine the properties of finite nuclei, and perform pretty well. These Skyrme models are also chosen such that they span over the present uncertainties in the EoS properties, e.g. slope of the symmetry energy $L_\sym$, etc. In addition, we provide the new EoS within the format of the CompOSE catalog~\cite{compose}, which offers an access to a large set of EoS ready to be implemented in astrophysical modeling. It also
allows us to analyze the recent progresses in the field of dense matter physics. 

The present paper is ordered as follow: in Section~\ref{sec:unifmatt} we compare the Skyrme and $\ChiEFT$ predictions in uniform matter and analyze their systematical differences, especially for low-density neutron matter. Then in Sec.~\ref{sec:cldm} we compare the various crust EoS provided by the CLDM based on different nuclear interactions. Finally, we show our results in Sec.~\ref{sec:results} using the same files and units as the ones uploaded on the CompOSE catalog.

\begin{table*}[t]
\centering
\caption{Nuclear empirical parameters for the Skryme models and the Hamiltonians derived from $\ChiEFT$ used in the present work. The lowest order NEP are defined at $n_\sat$ ($E$, $L$/$n$, $K$), while the higher order ones ($Q$, $Z$) are defined from a fit of SM and NM up to 10$n_\sat$ for Skyrme, and a fix to reach 2M$_\odot$ for the $\ChiEFT$ models. The last four columns show the low-density correction parameters $b_{\sat}$ and $b_{\sym}$, and the effective mass $m^*_\sat$ at saturation in symmetric matter and the effective mass splitting $\Delta m^*_\sat$, see text for more details. }
\label{tab:empirical:hamiltonians}
\tabcolsep=0.11cm
\def\arraystretch{1.6}
\begin{tabular}{lccccccccccccccc}
\hline\noalign{\smallskip}
Model & $E_\sat$ & $n_\sat$ & $K_\sat$ & $Q_\sat$& $Z_\sat$  & $E_\sym$ &  $L_\sym$ & $K_\sym$ & $Q_\sym$ & $Z_\sym$ & $b_\sat$ & $b_\sym$ & $m^*_\sat$ & $\Delta m^*_\sat$ \\
 &(MeV) & (fm$^{-3}$) &(MeV) & (MeV) & (MeV) & (MeV) & (MeV) & (MeV) & (MeV) & (MeV) & & & ($m_N$) & ($m_N$) \\
\hline\noalign{\smallskip}
H1$_\mm$&  -17.0 & 0.186 &  261 & -220 & -200  & 33.8 & 46.8 & -154 & 700 & 500 &  11.37 & 10.46 & 0.59  & 0.43  \\
H2$_\mm$ &  -15.8 & 0.176 & 237 & -220 & -200 & 32.0 & 43.9 & -144 & 700 & 500 & 9.59 & 9.34  & 0.61 & 0.41 \\
H3$_\mm$ &  -15.3 & 0.173 & 232 & -220 & -200 & 31.8 & 50.6 & -96 & 700 & 500 & 10.35 & 20.56  & 0.61 & 0.34 \\
H4$_\mm$&  -15.0 & 0.169 &  223 & -220 & -200 & 31.0 & 42.1 & -138 & 700 & 500 & 8.37 & 8.70  & 0.63 & 0.38 \\
H5$_\mm$&  -13.9 & 0.159 &  207 & -220 & -200 & 29.4 & 40.2 & -128 & 700 & 500 & 6.41 & 8.24  & 0.66 & 0.33 \\
H7$_\mm$ &  -13.2 & 0.139 & 201 & -220 & -200 & 28.1 & 36.5 & -150 & 700 & 500 &  9.40 &-1.46  & 0.67 & 0.41 \\
DHS$^{L59}_\mm$ &  -14.0 & 0.168 &  200 & -220 & -200 & 31.4 & 58.9 & -30 & 700 & 500 & 9.00 & 10.00  & 1.0 & 0.0 \\
DHS$^{L69}_\mm$ &  -14.6 & 0.173 & 216 & -220 & -200 & 33.7 & 69.0 & -20 & 700 & 500 & 9.00 & 10.00  & 1.0 & 0.0 \\
\noalign{\smallskip}\hline\noalign{\smallskip}
BSK14$_\mm$  & -15.9 & 0.159 & 239 & -88 &  -896 & 30.00 & 43.9 & -152 & 213 &  -676 &  1.22  &  0.01 &   0.80 &    0.03 \\
BSK16$_\mm$  & -16.1 & 0.159 & 242 & -91 &  -895 & 30.00 & 34.9 & -187 & 245 &  -696 &  1.22 &  0.01  &   0.80 &    0.04 \\
F0$_\mm$  & -16.0 & 0.162 & 230 & -124 &  -749 & 32.00 & 42.4 & -113 & 294  &  -625  &  1.23  &  0.06 &  0.69 &   -0.19  \\
LNS5$_\mm$  & -15.6 & 0.160 & 240 & -113 &  -574 & 29.15 & 50.9 & -119 & 195&  -853  &  1.26 &  0.00  & 0.60 &    0.23  \\
RATP$_\mm$  & -16.0 & 0.160 & 240 & -109 & -697 & 29.26 & 32.4 & -191 & 266  &  -923  &  1.25 & -0.00&  0.67 &    0.26\\
SGII$_\mm$  & -15.6 & 0.158 & 215 & -90 &  -874 & 26.83 & 37.6 & -146 & 221  &  -915 &  1.25  &  0.00 & 0.79 &    0.28 \\
SLY5$_\mm$  & -16.0 & 0.160 & 230 & -104 &  -749 & 32.03 & 48.3 & -112 & 195&  -326 &  1.26  & -0.01  &  0.70 &   -0.18\\
\noalign{\smallskip}\hline
\end{tabular}
\end{table*}

\begin{figure*}[t]
\centering
\includegraphics[scale=0.34]{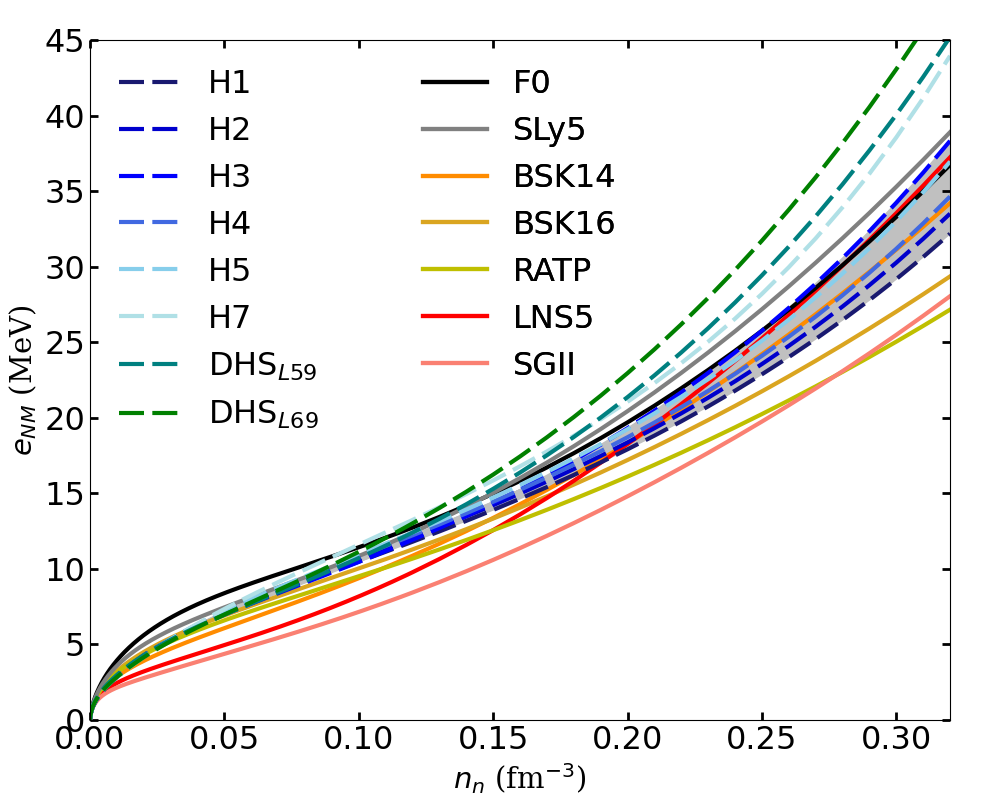}
\includegraphics[scale=0.34]{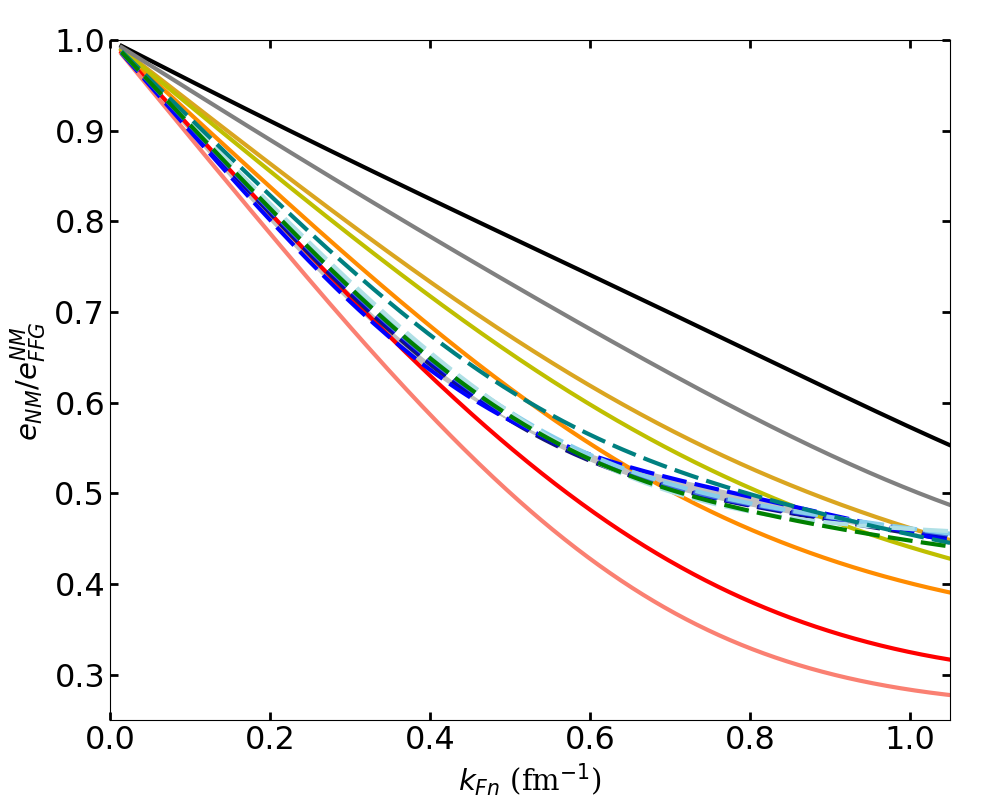}\\
\includegraphics[scale=0.34]{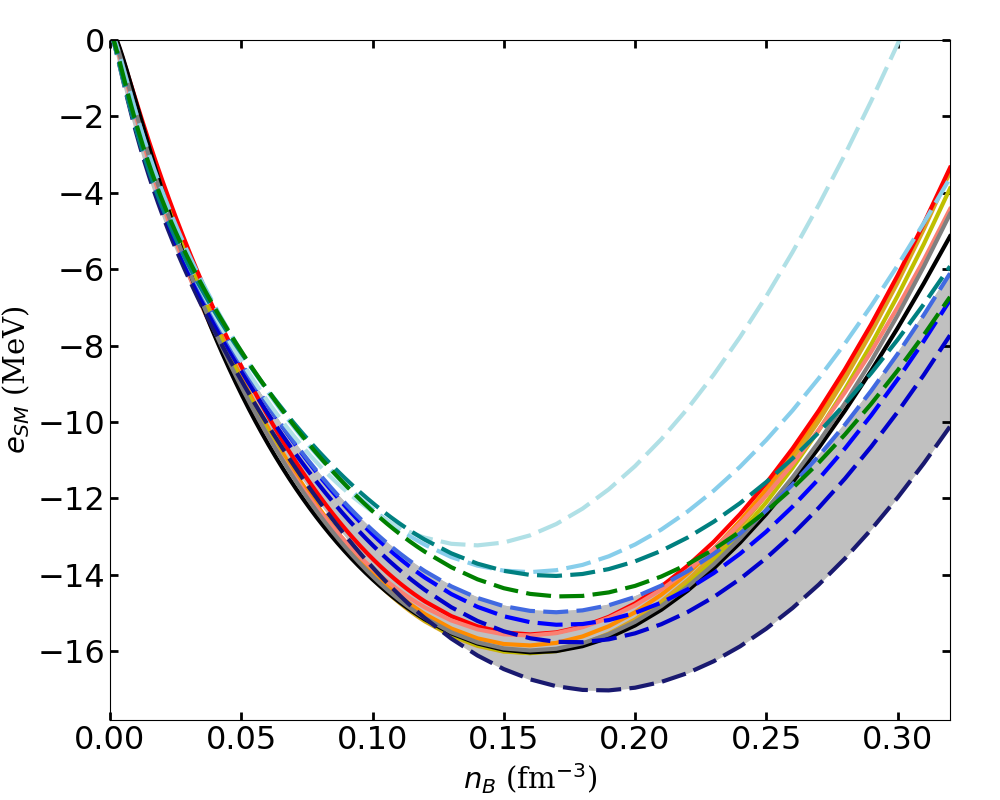}
\includegraphics[scale=0.34]{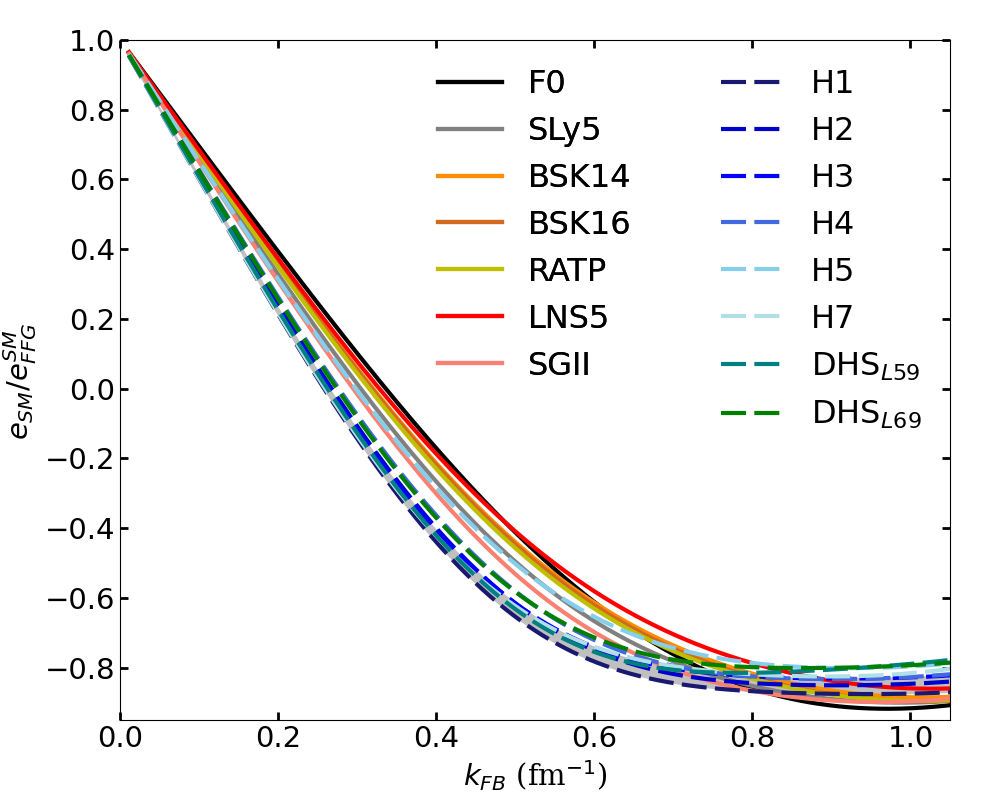}
\caption{Energy per particle in uniform matter (left), for NM (top) and for SM (bottom), normalized by the Free Fermi Gas energy $e_\ffg$ (right). Plots are drawn as function of the densities, $n_n$ in NM and $n_B$ in SM, as well as function of the Fermi energy, $k_{Fn}$ in NM and $k_{FB}$ in SM. Note that in SM $n_n=n_B/2$ and $k_{Fn}=k_{FB}$. We plot the 8 $\ChiEFT$ Hamiltonians (dashed lines), a gray band for H1-H4, and 7 Skyrme interactions (solid lines), see the legend for more details. Note that for the clarity of the legend in all the figures shown in this paper, we drop the index MM to all models. It is implicit that in all this work, we use the meta-model adjusted to the predictions of the microscopic interactions. For $\ChiEFT$ Hamiltonians however, the extrapolation at high density stands for only one among all possible realizations of the Hamiltonian guided by the condition to reproduce the radio observations of massive pulsars, as detailed in the text.}
\label{fig:enerUnif}
\end{figure*}

\begin{figure*}[t]
\centering
\includegraphics[scale=0.34]{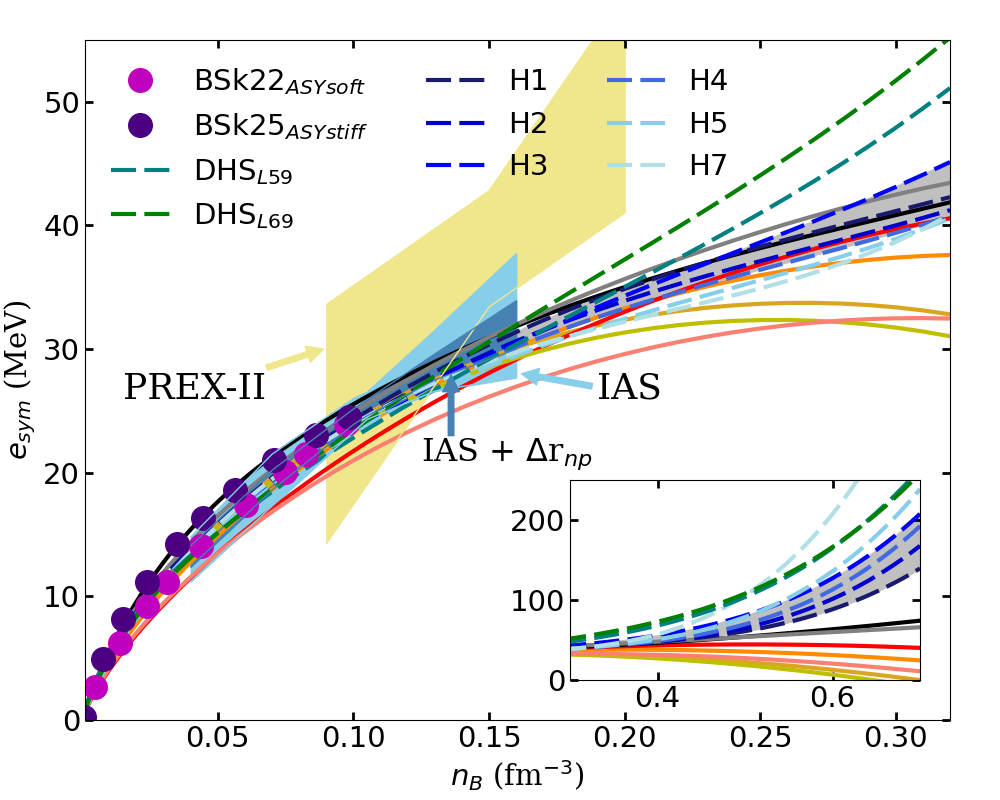}
\includegraphics[scale=0.34]{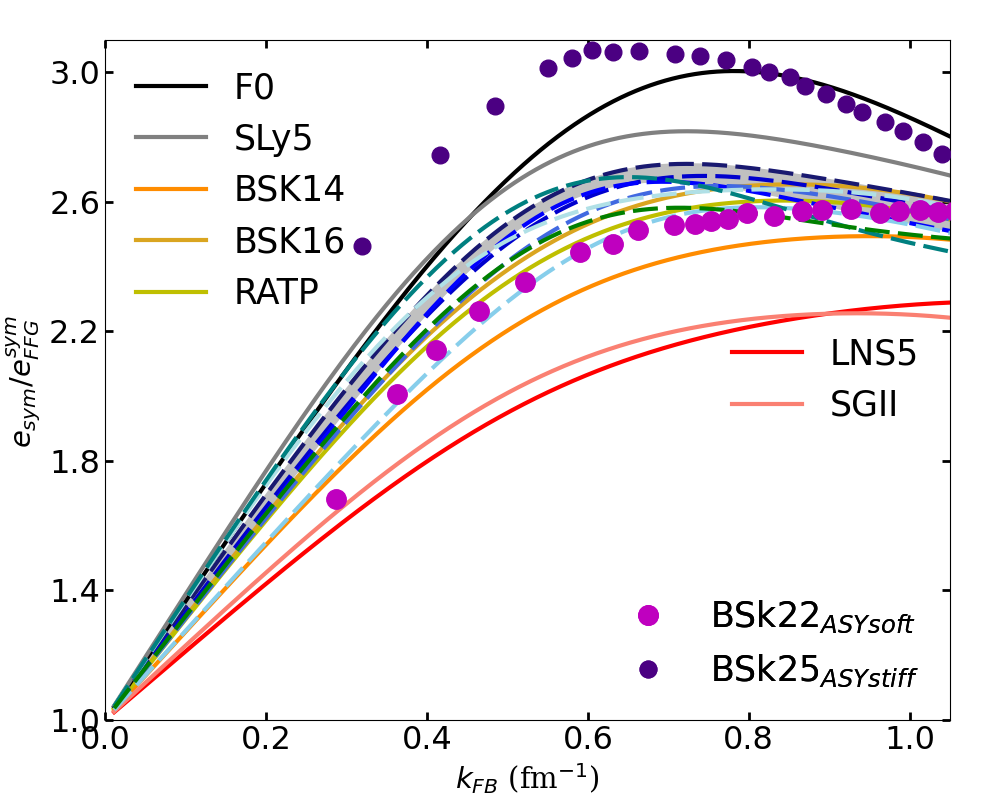}
\caption{Symmetry energy w.r.t to baryon density (left) and normalized by the FFG symmetry energy w.r.t Fermi momentum (right). Inset on left plot shows $e_{sym}$ at high densities. Continuous (dashed) lines shows the meta-model fitted to Skyrme ($\ChiEFT$) models. Dark purple and magenta dots show predictions from Brussels-Montreal microscopic models BSk22 and BSk25~\cite{Pearson18}. Light (darl) blue bands show constraints from isobaric analog state IAS (IAS + neutron skin)~\cite{Danielewicz2014}. The recent constraint from neutron skin experiments by PREX-II is shown in yellow band \cite{Brendan2021}.}
\label{fig:enerSym}
\end{figure*}

\begin{figure}[t]
\centering
\includegraphics[scale=0.34]{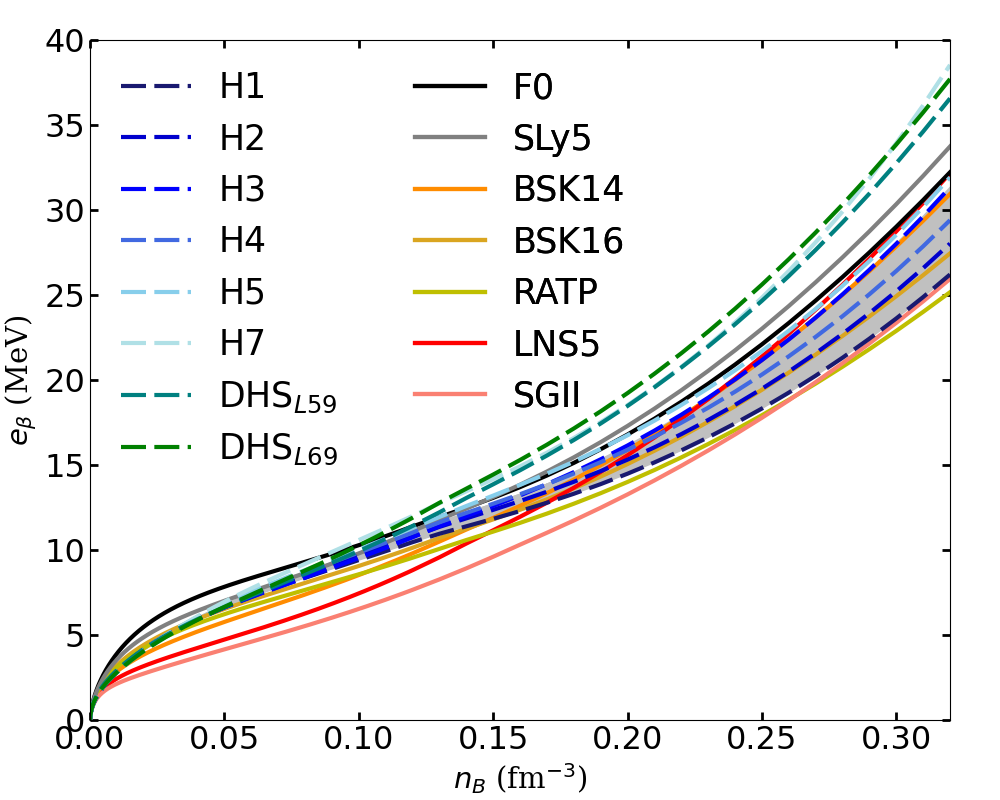}
\caption{Uniform matter energy in $\beta$-equilibrium.}
\label{fig:enerBetaUnif}
\end{figure}

\section{Uniform matter in the core of neutron stars}
\label{sec:unifmatt}

The core of NS is composed of uniform matter at $\beta$-equilibrium, whose properties are strongly related to the symmetry energy originating from the energy difference between neutron matter (NM) and  symmetric matter (SM). We therefore first discuss uniform matter properties for the set of models considered in this analysis.
Our description of uniform matter is based on the meta-model (MM)~\cite{Margueron2018a} which is calibrated on existing models, such as the Skyrme model or the $\ChiEFT$ Hamiltonian, by the use of the nuclear empirical parameters (NEPs). These NEPs are obtained from the derivatives of the energy per particle in SM and the symmetry energy, as detailed in Ref.~\cite{Margueron2018a} for instance. 

The reproduction of the Skyrme predictions for uniform matter is presented in Ref.~\cite{Margueron2018a} and in Ref.~\cite{Grams:2021b} we have presented the calibration of the MM to the many-body perturbation theory predictions based on the  $\ChiEFT$ Hamiltonians: 
H1-H7~\cite{Drischler2016} and the DHS$_{L59}$-DHS$_{L69}$~\cite{Drischler2021}.
Note that in the present work, the Skyrme MM adjustment is slightly different from the one shown in Ref.~\cite{Margueron2018a} where it was analytically fixed to reproduce Skyrme's predictions ($E/A$ and $P$) at 4$n_\sat$. We replaced this -- somehow arbitrary -- prescription by a fit over the densities in SM and NM and from $n_\sat$ up to $10 n_\sat$ taking the parameters $Q_{\sat/\sym}$ and $Z_{\sat/\sym}$ are free parameters, the other NEPs being taken at their predicted values. This new prescription avoids extrapolating the MM in dense region where it may deviate from the original Skyrme's predictions.

In the following, we investigate the following nucleonic models:
\begin{itemize}
\item Skyrme forces: we select several Skyrme forces which have been adjusted to the ground-state properties of finite nuclei, and therefore are expected to reproduce well SM around $n_\sat$. Their predictions in NM however differ largely and they represent the actual uncertainty for NM EOS based on phenomenological approaches. In the following, we select BSK14\cite{Goriely07}, BSK16\cite{Chamel08}, F0\cite{Lesinski06}, LNS5\cite{Cao06}, RATP\cite{Rayet82}, SGII\cite{Nguyen81}, SLy5\cite{Chabanat98}. Anticipating further results, the wide predictions in NM from these approaches mostly reflect their different predictions for the parameter $L_\sym$, see Table~\ref{tab:empirical:hamiltonians}.
\item $\ChiEFT$: this approach takes as experimental constraints the NN scattering properties in vacuum complemented with the binding energy in $d$ and $^3$He. We have selected several of the latest predictions in SM and NM which are H1-H7~\cite{Drischler2016} (except H6), and DHS$_{L59}$-DHS$_{L69}$ \cite{Drischler2021}. These eight Hamiltonians explore uncertainties in the chiral NN and 3N interactions. Since the MM needs constraints at high density, beyond the break-down density of these approaches, we have additionally fixed the high order empirical parameters $Q_{\sat/\sym}$ and $Z_{\sat/\sym}$ such that these EOS predict a maximal mass $M_\tov$ above the largest observed mass at about $2M_\odot$. Since this prescription is added on top of the original $\ChiEFT$ prediction, we decided to fix the high-order NEPs to the same values for all EOS.
\end{itemize}

The NEPs predicted by these models are given in Table~\ref{tab:empirical:hamiltonians}. Note that the third and forth order parameters ($Q$ and $Z$) are not fixed at saturation density, but rather imposed by a fit over the densities (for Skyrme) or by the requirement to reach 2M$_\odot$ (for $\ChiEFT$).

The predictions in SM and NM are shown in Fig.~\ref{fig:enerUnif}, see caption for details. A gray band captures the uncertainties originating from $\ChiEFT$ Hamiltonians H1-H4, which reproduce experimental nuclear masses at best, see table~\ref{table:nucleifit}. In NM the gray $\ChiEFT$ band is much narrower than the dispersion among the Skyrme models, see top panels of Fig.~\ref{fig:enerUnif}. The reason is related to the fact that $\ChiEFT$ theory is well-suited to describe low-density NM, which is directly constrained by the nucleon-nucleon phase shifts and three-body forces. Skyrme interactions are calibrated using the properties of finite nuclei, which reflect more directly the properties of SM, as it can be seen from Fig.~\ref{fig:enerUnif}(bottom panels), where the $\ChiEFT$ band is now larger than the dispersion among the Skyrme models. At first sight, the predictions from $\ChiEFT$ and Skyrme models are complementary: the former describes NM better, while the latter is better for SM.

The low-density energy per particle is shown in the right panels of Fig.~\ref{fig:enerUnif}, as function of the Fermi momentum,
$k_{Fn}=(3\pi^2 n_n)^{1/3}$ (NM)
and $k_{FB}=(3\pi^2 n_B/2)^{1/3}$ (SM). The Free Fermi Gas (FFG) energy per particle $e_\ffg^\mathrm{NM}=3\hbar^2k_{F}^2/(10m_N)$, where $m_N$ is the nucleon mass, $k_F = k_{Fn} $ in NM and $k_F = k_{FB}$ in SM, scales the energies shown in the left panels of Fig.~\ref{fig:enerUnif}.
In this representation, it is clear than the predictions from Skyrme models appear to be almost unconstrained in NM, as already suggested in Ref.~\cite{Roggero:2015}, while the ones based on $\ChiEFT$ are very consistent among each other. In a previous analysis~\cite{Grams:2021a}, we have compared in detail the $\ChiEFT$ predictions and the SLy4 Skyrme model, which is almost identical to the SLy5 presented here. By investigating other Skyrme models, which perform rather well for the ground state of finite nuclei, we now explore more widely the Skyrme's uncertainties in the predictions for the properties of low-density NM. These properties are interesting since low-density NM is close to the unitary gas limit~\cite{Carlson:2008,Bulgac:2012,Vidana:2021}, which is universal from nuclear systems to cold atom gas~\cite{Navon:2010}. From our previous analysis, we have concluded that the crust composition ($A_\cl$, $Z_\cl$) depends mainly on the properties of SM, while NM influences mostly the energy per particle, the pressure and the sound-speed in non-uniform matter. We will use these results in the discussion of the crust prediction in the next section. 

The symmetry energy, defined as,
\begin{equation}
e_\sym(n) = e_\nm(n) - e_\sm(n) \, ,
\end{equation}
where $e=E/A$, is shown in Fig.~\ref{fig:enerSym}. It represents the cost in energy per particle to convert SM into NM. In dense matter at $\beta$-equilibrium, this energy is provided by electrons at their Fermi energy. Despite the dispersion of $\ChiEFT$ predictions in SM, these models predict  symmetry energy in a narrower band compared to the Skyrme's one, since Skyrme models are penalized by their poor reproduction of NM.
We additionally display constraints from nuclear experiments: the light (dark) blue band are shown the constraints from isobaric analog state IAS (IAS + $\Delta r_{np}$, i.e., IAS + neutron skin)~\cite{Danielewicz2014}, and the yellow band shows the PREX-II predictions from the neutron skin in Pb~\cite{Brendan2021}, leading to $E_\sym = 38.1 \pm 4.7$ MeV and $L_\sym = 106 \pm 37$ MeV. We note that all models predict symmetry energy lower than PREX-II band when $n_B \approx n_{\sat}$ and inside the very lower band for $n_B \approx 0.1$ fm$^{-3}$, which shows a disagreement between PREX-II measure and the model predictions. Regarding the constraints from IAS~\cite{Danielewicz2014} all $\ChiEFT$ models predict symmetry energy located inside both blue bands while most Skyrme models predict $e_\sym$ inside the light blue band, except SGII that is always softer and LNS5 which crosses the constraint band just around saturation. For the more stringent constraint, IAS + neutron skin, we note that SGII and LNS5 predict low values for the symmetry energy. RATP while compatible with the symmetry energy band around saturation density tends to be as soft as SGII and LNS5 above saturation density.
Similarly to Fig.~\ref{fig:enerUnif}, we show in Fig.~\ref{fig:enerSym} (right panel) the symmetry energy normalized by the FFG prediction as function of the nucleon Fermi momentum. It is interesting to remark the very large deviation from one Skyrme model to the other on the vertical axis, which reflects the influence of the interaction to the symmetry energy (potential and effective mass contributions), see discussion in Ref.~\cite{Somasundaram2021} for instance. At low density, for instance $k_{FB}=0.4~\fmi$ ($n_B\approx 0.004~\fmiq$), the potential term contributes to double the symmetry energy. The potential term in $\ChiEFT$ models increases the normalized symmetry energy up to $k_{FB}=0.6$-$0.7~\fmi$ ($n_B\approx 0.015~\fmiq$), reaching $2.5$-$2.7$, and then it decreases slowly as the Fermi momentum $k_{FB}$ increases. This low-density behavior is however not characteristic of all Skyrme models: Some Skyrme interactions predict the symmetry energy to be lower than the $\ChiEFT$ predictions (BSK14, LNS5,  SGII) while some others predict it to be larger than the $\ChiEFT$ predictions (F0, SLy5).
In Fig.~\ref{fig:enerSym} we also show symmetry energy predictions from BSk22 and BSk25 Skyrme interactions~\cite{Pearson18}. These two models represent the two extreme cases investigated in Ref.~\cite{Pearson18}: BSk22 is ASYsoft (BSk25 is ASYstiff) below saturation density, as shown in Fig.~\ref{fig:enerSym}. The symmetry energy from BSk22 is similar to the $\ChiEFT$ band, while BSk25 predicts a large value for the symmetry energy below saturation density. In Ref.~\cite{Pearson18}, these two models are used in Hartree-Fock-Bologiubov modeling of the NS crust, as we will show in the following.

At high density, as shown in the left panel of Fig.~\ref{fig:enerSym}, several Skyrme models also predict a bending down of the symmetry energy (BSK14, BSK16, RATP, SGII, LNS5) that the $\ChiEFT$ do not predict.
The inset in Fig.~\ref{fig:enerSym} shows that the bending down of the symmetry energy around saturation density of these Skyrme models leads to a crossing of the zero axis at high density. This is a well known feature of phenomenological nuclear interactions such as Skyrme, see for instance Ref.~\cite{Danielewicz:2009} and references therein.

To conclude this section, we show in Fig.~\ref{fig:enerBetaUnif} the energy per particle at $\beta$-equilibrium in uniform matter. The condition to fulfill beta-equilibrium (without neutrinos and without muons) is the following: $\mu_n-\mu_p=\mu_e$. In cold catalyzed neutron stars, neutrinos do not contribute to the chemical equilibrium. Muons can however appear when $\mu_\mu\ge m_\mu c^2$ and they contribute to $\beta$-equilibrium through the thermodynamical relation: $\mu_e=\mu_\mu$. Charge neutrality is also imposed: $n_e+n_\mu=n_p$. These three conditions complemented by the baryon conservation number $n_B=n_n+n_p$ lead to uniquely determine the composition of npe$\mu$ matter. Below saturation density, the dispersion among the Skyrme predictions observed in Fig.~\ref{fig:enerBetaUnif} is very large, much larger than the predictions from $\ChiEFT$. In the following, we will investigate the impact of this large dispersion among the Skyrme predictions on the crust properties in non-uniform matter. This will be performed in the next section based on the CLDM approach.

Above saturation density and as the density increases, the $\ChiEFT$ band for the symmetry energy gets larger and larger. While some Skyrme models are close but out of the band, e.g., SLy5 (above) and RATP (bellow), the $\ChiEFT$ band represents a far estimation of the current uncertainties up to twice saturation density.

\section{The crust of neutron stars}
\label{sec:cldm}

In a recent paper~\cite{Grams:2021b}, we have presented the CLDM used in the present study for non-uniform matter and we have detailed how the bulk terms are connected to uniform matter by using the meta-model approach~\cite{Margueron2018a}, which reproduces uniform matter from $\ChiEFT$ and Skyrme approaches. The CLDM employed here describes isolated finite nuclei as well as nuclear clusters in the crust of NS. At variance with some other approaches where the experimental nuclear masses, when they exist, are included directly in the description of nuclear cluster energetic, we use in our approach the comparison to the experimental data to quantify the goodness of the models. This approach satisfies the unified prescription where finite systems and infinite nuclear matter are based on the same model.  For this reason, we first discuss in this section the constraints from experimental nuclear data, and then we present our crust EoS.

\subsection{Constraints from experimental nuclear data}

The CLDM defines the finite nuclei energy as $E_\nuc=E_\bulk+E_\fs$, which sums the bulk and finite-size (FS) contributions. The present notations are consistent with Ref.~\cite{Grams:2021b}, where more details are given.

\begin{table}[t]
\begin{center}
\caption{Standard parameters the surface and curvature terms employed in the CLDM approach considered in this work. These values are obtained from  averaging over the parameters given in Ref.~\cite{Carreau2019b}.
}
\label{table:stdParam}
\tabcolsep=0.25cm
\def\arraystretch{1.5}
\begin{tabular}{ccccc}
\hline\noalign{\smallskip}
$\sigma_\mathrm{surf,sat}^\std$ & $\sigma_\mathrm{surf,sym}^\std$ & $p_\surf ^\std$ & $\sigma_\mathrm{curv,sat}^\std$ &  $\beta_\mathrm{curv}^\std$ \\
MeV~fm$^{-2}$ & MeV~fm$^{-2}$ & & MeV~fm$^{-1}$ & \\
\noalign{\smallskip}\hline\noalign{\smallskip}
1.1 & 2.3$^1$ & 3.0 & 0.1 & 0.7 \\
\noalign{\smallskip}\hline
\end{tabular}
\end{center}
$^1$ Note the associated value $b_\surf=29.9$.
\end{table}

For each nucleus defined by its mass $A$ and its charge $Z$ localizing it over the nuclide chart, the mechanical equilibrium is imposed: $P_\nuc = n_\cl^2 \partial e_\nuc / \partial n_\cl = 0$, where $n_\cl$ is the nucleus density. This variational condition fixes the nuclear cluster density $n_\cl(A,Z)$ at equilibrium: each nuclei have a different density, and they overall explore the density dependence of the CLDM model around saturation density, at variance with the liquid-drop model where the model coefficients are kept constant. 

The bulk term, assuming the local density approximation, originates from uniform matter calculations presented in Sec.~\ref{sec:unifmatt}.
The FS term is composed of the Coulomb, surface and curvature contributions, as described in the FS4 prescription in Ref.~\cite{Grams:2021b}. These terms are moreover fine-tuned to better reproduce the experimental binding energies for $A\ge 12$ and $Z\ge 6$, as detailed in Ref.~\cite{Grams:2021b}. The standard parameters for the surface and curvature terms are given in Tab.~\ref{table:stdParam}. In the practical implementation of the CLDM, each of these values are multiplied by the parameters $\mathcal{C}_i$ ($\mathcal{C}_{\surf, \sat}$, $\mathcal{C}_{\surf, \sym}$, $\mathcal{C}_{\curv, \sat}$, $\mathcal{C}_{\beta}$) as,
\begin{eqnarray}
\sigma_{\surf, \sat} &=& \mathcal{C}_{\surf, \sat} \sigma_{\surf, \sat}^{\std}\, ,\\
\sigma_{\surf, \sym} &=& \mathcal{C}_{\surf, \sym} \sigma_{\surf,\sym}^\std\, , \\ 
\sigma_{\curv, \sat} &=& \mathcal{C}_{\curv, \sat} \sigma_{\curv, \sat}^{\std}\, , \\
\beta_\mathrm{curv} &=& \mathcal{C}_\beta \beta_\mathrm{curv}^{\std} \, ,
\end{eqnarray}
and the Coulomb energy is corrected with the coefficient $\mathcal{C}_\coul$.
Since these coefficients incorporate in an effective way the slight correction due to the actual density profile in finite nuclei, which we approximate by a hard sphere in the CLDM, the values for these variational parameters $\mathcal{C}_i$ are expected to be close to 1 as shown in Table~\ref{table:nucleifit}.

\begin{table*}[t]
\centering
\caption{Optimization of the parameters $\mathcal{C}_i$ over the nuclide chart, for $A\ge 12$ and $Z\ge 6$, and considering the experimental energies from the 2016 Atomic Mass Evaluation (AME)~\cite{AME2016}
for the interactions considered in this work (Skyrme and $\ChiEFT$).
Odd-even mass staggering parameters $\Delta_\sat$ and $\Delta_\sym$,
$\chi_E$ for the optimization with (without) odd-even mass staggering.
In the two last columns, the minimal energy $e_\cl$ for each model is given together with its position in the nuclide chart $(A,Z)_{\min}$. Experimental binding energy are  $e(^{56}$Fe$)$ = -8.79 MeV and  $e(^{68}$Zn$)$ = -8.76 MeV.}
\label{table:nucleifit}
\tabcolsep=0.2cm
\def\arraystretch{1.5}
\begin{tabular}{ccccccccccc}
\hline\noalign{\smallskip}
Model &  $\mathcal{C}_\coul$ & $\mathcal{C}_{\surf, \sat}$ & $\mathcal{C}_{\surf, \sym}$ & $\mathcal{C}_{\curv, \sat}$ &  $\mathcal{C}_{\beta}$ & $\Delta_\sat$ & $\Delta_\sym$ & $\chi_E$ & $\min(e_\nuc)$ & (A,Z)$_{\min}$ \\
      &    &    &    &    &    & (MeV) & (MeV) & (MeV)& (MeV) & \\
\hline\noalign{\smallskip}
BSK14$_\mm$ &   0.965 &   1.002 &   0.890 &   0.770 &   1.100 & 5.75  & -7.35 & 2.61 (2.74)  & -8.72 &   (68,30)\\
BSK16$_\mm$ &   0.973 &   1.076 &   0.876 &   0.777 &   0.817 & 5.68  & -4.88 &   2.66 (2.79)  & -8.70 &   (68,30)\\
LNS5$_\mm$  &   0.949 &   0.916 &   0.668 &   0.531 &   1.646 & 5.69 & -4.84 &   2.66 (2.79) &  -8.75 &   (68,30)\\
RATP$_\mm$ &   0.970 &   1.095 &   0.676 &   0.583 &   0.694 & 5.54  & 0.22 &   2.81 (2.93)  &  -8.70  &   (68,30)\\
SGII$_\mm$ &   0.952 &   0.942 &   0.406 &   0.262 &   1.981 & 5.42 & 4.97 &   2.95 (3.07) & -8.75 &   (68,30)\\
F0$_\mm$ &  0.966 &  1.064 &  1.264 &  1.144 &  0.922 &  5.82 & -10.47 &   2.60 (2.73) &  -8.65 &   (68,30)\\
SLy5$_\mm$ &   0.967 &   1.039 &   1.257 &   1.116 &   0.980 & 5.85 & -11.26 &   2.59 (2.72) & -8.67 &   (68,30) \\
\noalign{\smallskip}\hline\noalign{\smallskip}
H1$_\mm$ &   0.966 &   1.582 &   1.307 &   0.942 &  -0.255 & 5.53 & 0.22 &   3.07 (3.18) &  -8.17 &   (68,30)\\
H2$_\mm$ &   0.927 &   1.037 &   1.123 &   1.053 &   1.141 & 5.79 & -9.46 &   2.58 (2.71) &  -8.37 &   (70,30)\\
H3$_\mm$ &   0.913 &   0.847 &   1.095 &   1.127 &   1.560 & 5.94 & -14.83 &   2.65 (2.77) & -8.61 &   (68,30)\\
H4$_\mm$ &   0.903 &   0.716 &   0.967 &   1.144 &   1.833 & 5.99 & -16.85 &   2.77 (2.89)  &  -8.84 &   (68,30)\\
H5$_\mm$ &   0.868 &   0.325 &   0.646 &   1.215 &   2.616 & 6.32 & -29.26  &   3.78 (3.86) &  -9.83 &   (58,26)\\
H7$_\mm$ &   0.858 &   0.110 &   0.339 &   1.086 &   3.235 & 6.61 & -43.69 &   5.32 (4.95) & -10.23 &   (56,26)\\
DHS$_{L59,\mm}$ &   0.853 &   0.370 &   1.038 &   1.686 &   2.089 & 6.41 & -33.40 &   4.34 (4.41) & -9.66  &   (58,26)\\
DHS$_{L69,\mm}$ &   0.871 &   0.567 &   1.571 &   1.971 &   1.649 & 6.28 & -28.51 &   4.02 (4.10) & -9.10 &   (60,26)\\
\noalign{\smallskip}\hline
\end{tabular}
\end{table*}

\begin{figure}[t]
\centering
\includegraphics[scale=0.34]{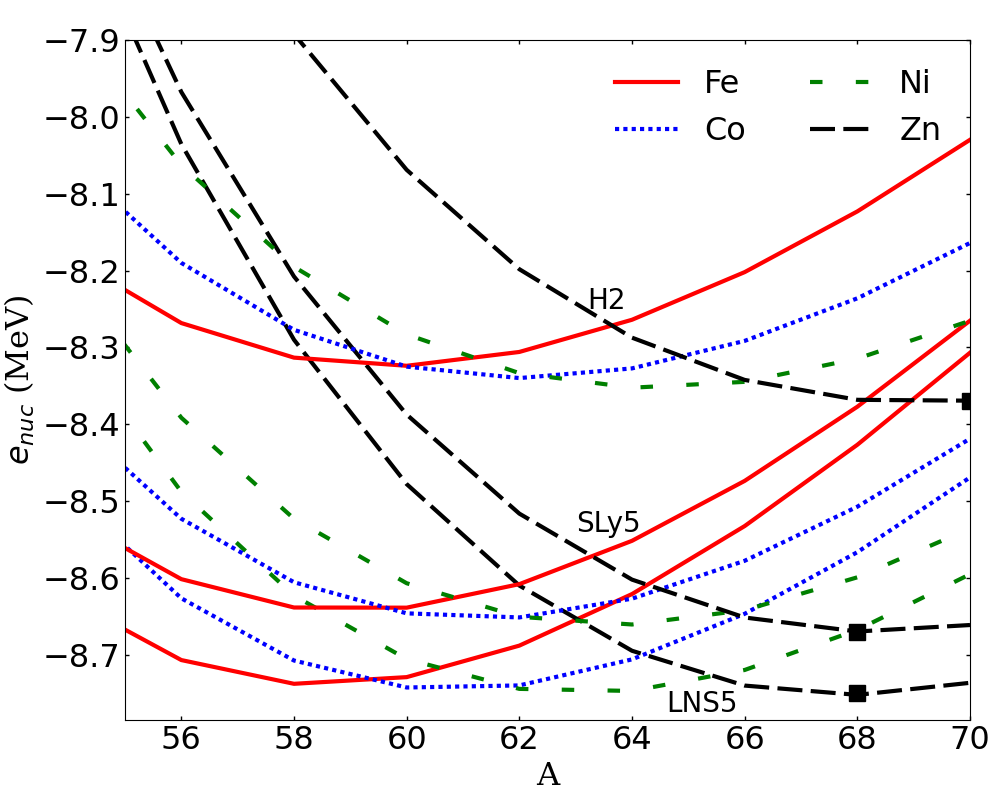}
\caption{Binding energy for the Iron group using the CLDM with SLy5, LNS5 and H2. A black square marks the lowest energy for each model.}
\label{fig:minEnerIrongroup}
\end{figure}

A small improvement of the fit to the experimental nuclear masses has been implemented in the present work in order the reduce the effect of the odd-even mass staggering in the data: we have corrected the experimental masses as, $\tilde{E}_\ex^i = E_\ex^i - \Delta E_\ex^i$, with
\begin{equation}
\Delta E_\ex^i = \left[\Delta_\sat + \Delta_\sym \left(\frac{N_i-Z_i}{A_i}\right)^2 \right] \, A_i^{-1/3} \, \delta(N,Z) \, .
\end{equation}
where $\delta(N,Z) = 1$ if $N$ and $Z$ are odd, $0$ if either $N$ or $Z$ is odd, and $-1$ if both $N$ and $Z$ are even~\cite{BohrMottelson1969}. The parameters $\Delta_\sat$ and $\Delta_\sym$ are varied together with the CLDM parameters $\mathcal{C}_i$ in the fit to the experimental masses. The loss function $\chi_E$, required for the fit, is defined as,
\begin{eqnarray}
\chi_E &=& \left[ \frac{1}{N} \sum_{i=1}^{N} ( \tilde{E}_\ex^i - E_\nuc^i )^2 \right]^{1/2} \, ,
\label{eq:chiE}
\end{eqnarray}
where $N=3375$ (2443 without extrapolated data) is the number of considered nuclei from the experimental nuclide chart and $E_\nuc^i$ the energy from the CLDM model for the nucleus $i$.
By comparing $\chi_E$ and $\chi_{E/A}$ (defined as function of $E/A$ instead of $E$), the impact of the loss function has been shown to be non-negligible in general, but smaller than other uncertainties~\cite{Grams:2021b} such as for instance the one coming from varying the model for the bulk. We will thus employ only $\chi_E$ for the fit, without considering the uncertainties originating from a different measure of the goodness of the model to reproduce experimental nuclear masses. Note that we also checked that our predictions are weakly impacted by incorporating the extrapolated data for the experimental masses, compared to the case where we would restrict the nuclear chat only to the measured masses.

The results of the fit to finite nuclei are shown in Table~\ref{table:nucleifit}. The coefficients $\mathcal{C}_i$ are in general of the order of 1, as expected.
The odd-even mass staggering parameters $\Delta_\sat$ and $\Delta_\sym$ are also shown. The isoscalar parameter $\Delta_\sat$ is similar to the value suggested in Ref.~\cite{Vogel:1984} (7.2~MeV). The isovector parameter $\Delta_\sym$ however is much smaller than the one suggested in Ref.~\cite{Vogel:1984} (-44~MeV), where the fit is done by considering only $A > 40$ nuclei.
We indeed found that this quantity is very sensitive to the nuclei which we fit it. When we select only $A > 40$ nuclei $\Delta_\sym$ change from -7.35 (-11.26) to $\approx$-16.83 (-16.60) MeV for BSK14 (SLy5). The value of $\Delta_\sym$ therefore varies from one region of the nuclear chat to another. It is difficult to determine a constant value over the nuclide chart, but its influence is small since nuclei do not explore large isospin asymmetries ($N<2Z$ for heavy nuclei).

With a loss function $\chi_E \lesssim 3$~MeV, the $\ChiEFT$ models H1 to H4 are the more accurate ones to reproduce the experimental masses. They are comparable with the Skyrme models considered here. Note that our conclusion remains valid even when we do not correct for the odd-even mass staggering in the nuclear data, see number is parenthesis in the $\chi_E$ column on Table \ref{table:nucleifit}. The nuclear masses however disfavor the $\ChiEFT$ models: H5, H7, DHS$_{L59}$ and DHS$_{L69}$. In the following, Hamiltonians H1-H4 will be considered as our best models and will be marked with a gray band, since they reproduce nuclear masses at best.

In the two last columns of table~\ref{table:nucleifit} are shown the minimal energy per particle obtained for each model. This quantity is sometimes used to define the offset of the energy per particle in the crust, see next sub-section. Experimentally, the nucleus $^{56}$Fe minimizes the energy, but it is a bit model dependent due to the approximation scheme. The search for the minimal energy per particle configuration is illustrated in Fig.~\ref{fig:minEnerIrongroup} for a few models (LNS5, SLy5 and H2) and for iron group isotopic chains (Fe, Co, Ni, Zn). For all the Skyrme models, the lowest energy configuration is $^{68}$Ni. This is also the lowest energy configuration for H1, H3, H4, while H2 prefers $^{70}$Ni and the other $\ChiEFT$ models prefer $^{56-60}$Fe nuclei.

\subsection{Crust equation of state}

The non-uniform matter in the crust is modeled according to the CLDM, which sums the nuclear cluster energy $E_\cl$, with bulk, FS and electron interaction contributions, the neutron fluid contribution $E_n$ as well as the kinetic electron gas $E_e$:
\begin{equation}
E_\tot = E_\cl + E_n + E_e \, .
\end{equation}
Note that as the electron density goes to zero, $n_e\rightarrow 0$, then $E_\cl\rightarrow E_\nuc$, previously defined.
Details on these different terms are given in Ref.~\cite{Grams:2021b}.

\begin{figure*}[t]
\centering
\includegraphics[scale=0.37]{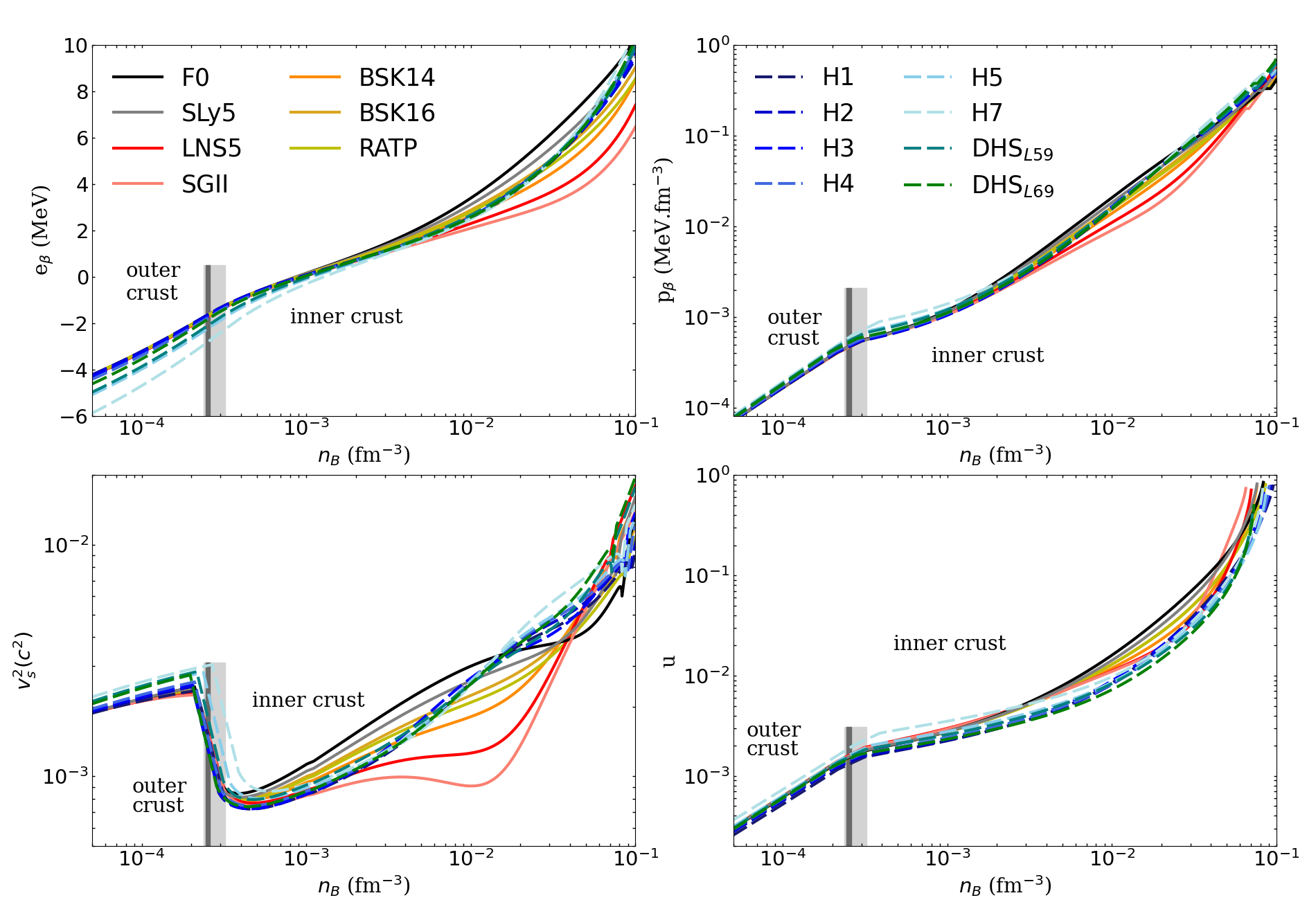}
\caption{Energy per particle, pressure, sound speed and volume fraction at beta equilibrium in the NS crust. Dark gray band shows the prediction for the outer-inner crust transition for the models that better reproduce nuclear masses (all Skyrme and H1-H4). Light gray band shows the uncertainty including all models.}
\label{fig:EOS}
\end{figure*}

The energy, pressure, sound speed and volume fraction, $u=V_\cl/V_{\rm WS}$ where $V_\cl$ is the cluster volume and $V_{\rm WS}$ the Wigner-Seitz cell volume, at $\beta$-equilibrium are shown in Fig.~\ref{fig:EOS} for a range of densities spanning over the outer (partially) and inner crust. As already suggest in Ref.~\cite{Grams:2021a}, these quantities are sensitive to the properties of low-density NM, as shown in Fig.~\ref{fig:enerUnif}.
The models LNS5 and SGII (F0 and SLy5), which predict binding energies in low-density NM lower (higher) than the $\ChiEFT$ band, also predict that the binding energy at $\beta$-equilibrium is lower (higher) in the density range going from $3~10^{-3}$ to $10^{-1}$~$\fmiq$.
Same remarks can be done for the pressure, the sound speed and the volume fraction at $\beta$-equilibrium. Vertical dark gray bands in Fig.~\ref{fig:EOS} shows the prediction for the outer-inner crust transition for the models that better reproduce nuclear masses, i.e., all Skyrme models and H1-H4, while light gray band shows the prediction including all models used in this work. Note the importance of reproducing experimental nuclear masses to better constraint the location of the neutron drip. 

It can also be remarked that in the outer crust the $\ChiEFT$ Hamiltonians are more spread than the Skyrme models. Since the Skyrme models reproduce better the experimental nuclear binding, they lead to tighter predictions in the outer crust. Note that the gray band localizing the predictions from H1-H4 $\ChiEFT$ Hamiltonians are in full agreement with the Skyrme's ones. One can thus conclude that the experimental nuclear masses are important constraints to accurately predict the EoS in NS outer crust. In the inner crust however, since a large amount of neutrons drip off clusters forming a neutron fluid, some properties such as the binding energy, the pressure, the sound speed and the volume fraction are largely impacted by the properties of uniform NM.

The transition from non-homogeneous matter (crust) to homogeneous matter (core) is obtained by computing the energy per particle for the two phases and comparing their values with the increase of the density. The lower energy is the favorable state. In the bottom-right panel of Fig. \ref{fig:EOS} the end of the volume fraction curves shows this transition. The spread on the end of curves shows the uncertainty on the crust-core transition for the 15 models investigated in this work. In Tab. \ref{tab:ndripNccNmaxCausal} we add the numerical values for the outer-inner crust transition together with the crust-core for the different models.

\section{Tables of equation of state}
\label{sec:results}

The fifteen EoS presented in this paper are available on the CompOSE catalog~\cite{compose} and can be interfaced with the Lorene library\footnote{\url{https://lorene.obspm.fr/}}. We briefly describe here a few quantities given in these tables and we start with the enthalpy.

The enthalpy per particle is defined as 
\begin{equation}
h(n_B)=\langle m\rangle c^2+e_{\tot}+p_{\tot}/n_B\, ,
\label{eq:enthalpy}
\end{equation}
where the mass $\langle m\rangle=(n_n m_n + n_p m_p)/n_B$ for the baryon density $n_B=n_n+n_p$ contributes to the rest mass term in Eq.~\eqref{eq:enthalpy}.
From Eq.~\eqref{eq:enthalpy} one could define the enthalpy per unit mass $m_\refr$ as $h/m_\refr$, where the unit mass $m_\refr$ is an arbitrary quantity. It could be taken to be the atomic unit mass $m_u$ \footnote{$m_u=931.494028(23)$~MeV~c$^{-2}$}, or the lowest measured mass of $^{56}$Fe  \footnote{$m(^{56}$Fe$)=930.411790$~MeV~c$^{-2}$}, and sometimes, it is just taken to be the neutron mass \footnote{$m_n=939.565413(6)$~MeV~c$^{-2}$}. As long as the definition of the rest mass is well defined, all these choices are equivalent.

The equilibrium of static neutron stars is given by the solution of the TOV equations~\cite{Tolman1939,Oppenheimer1939}.
In some numerical resolution of these equations, see for instance Ref.~\cite{Lindblom1992} or the Lorene library, the EOSs are sampled according to the log enthalpy per unit mass defined as,
\begin{equation}
H\equiv \log\left(\frac{h}{m_\refr c^2}-1\right) \, .
\label{eq:Enthalpy}
\end{equation}
In this case, it is important to fix $m_\refr$ to be below the lowest possible mass such that the ratio $h/m_\refr$ is always above $1$, and the quantity given by Eq.~\eqref{eq:Enthalpy} always well-defined. A natural choice would be to take $m(^{56}$Fe$)$ since it is the most stable system in nature. However, the NS EOS is not experimentally measured but instead is based on a modeling of the experimental nuclear masses. Due to the approximation scheme in the modeling, it is not given that the model mass reproduces the experimental quantity. It shall then be checked that the choice for $m_\refr$ ensures the ratio $h/m_\refr$ is above 1. To address this question in a more quantitative way, we have investigated in Fig.~\ref{fig:minEnerIrongroup} the location of the minimal energy nucleus, and we found that it is not necessarily $^{56}$Fe for all EOSs. For each EOS, we have then reported the value of the binding energy corresponding to the minimal energy nucleus in Tab.~\ref{table:nucleifit}. The reference mass shall thus be fixed to be the minimum energy per particle predicted by the considered model. From Tab.~\ref{table:nucleifit} we can check that min$(e_\nuc)/m(^{56}$Fe$)c^2<1$ for all Skyrme model we considered, but some of the $\ChiEFT$ models, while not being our preferred ones, do not respect this condition. Employing these models in TOV solver using \eqref{eq:Enthalpy} should then be performed carefully, and it could be necessary to lower down the value of the reference mass for these cases.

\begin{figure}[t]
\centering
\includegraphics[scale=0.35]{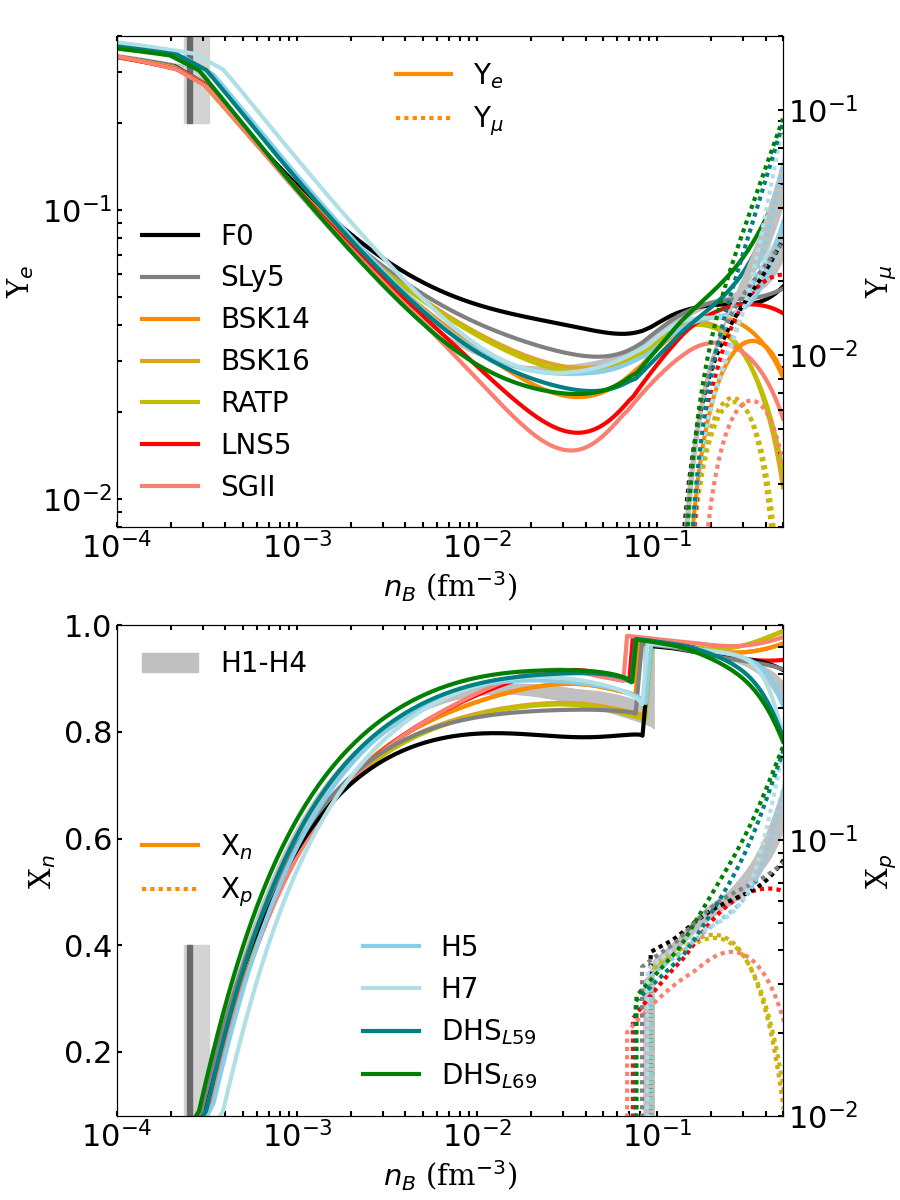}
\caption{Particle fractions. Top: leptons, $Y_e$ left axis ($Y_\mu$ right axis) in continuous (dotted) lines. Bottom: baryons, $X_n$ left axis ($X_p$ right axis) in continuous (dotted) lines.}
\label{fig:parFrac}
\end{figure}

The neutron, proton, electron and muon particle fractions are shown in Fig.~\ref{fig:parFrac}.
For densities lower than $10^{-3}~\fmiq$ all models agree well together, except H7, which slightly overestimates $Y_e$. At higher densities however, a dispersion among the model predictions appears.
Here also the low-density energy per particle in NM plays an important role: a reduction of the energy per particle in NM would ease the production of neutrons $\beta$-processes from electrons and protons, and thus would reduce the electron fraction $Y_e$. This is exactly what happens: the stiffest models, e.g. F0 and SLy5, predict the largest particle fractions in the density range going from $10^{-3}-10^{-1}~\fmiq$, while the softest ones, e.g. LNS5 and SGII, predict the lowest electron fractions.

Muons appear at around $0.1$-$0.2~\fmiq$, and this onset density is rather model independent (at least for density scale of the figure). The muon fraction increases for all models. At a few times saturation density, some models such as BSK14, BSK16, RATP, SGII and LNS5 predict however a bending down of the electron and muon density. This behavior results from the symmetry energy, see Fig. \ref{fig:enerSym}, which bends down to zero for these models as the density exceeds saturation density. As the symmetry energy reduces, it is more and more easy to produce neutrons and then the lepton fraction goes down to zero.

\begin{figure}[t]
\centering
\includegraphics[scale=0.37]{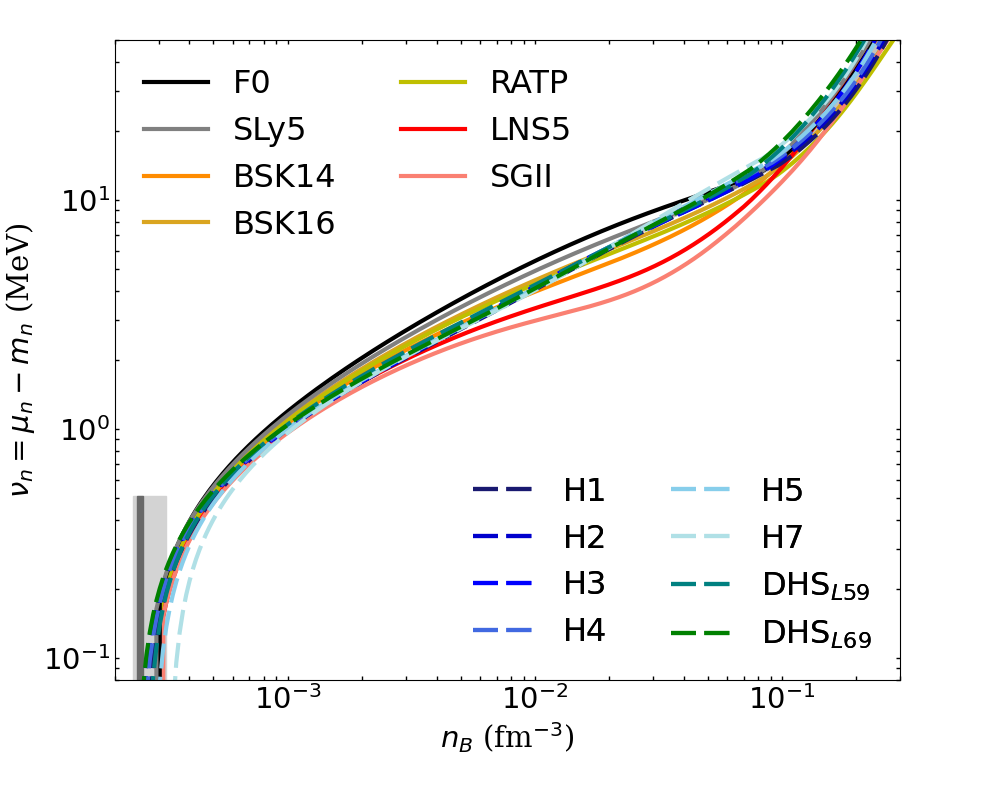}
\caption{Neutron chemical potentials in the inner crust with respect to the baryon density.}
\label{fig:chempot}
\end{figure}

The neutron chemical potential is shown in Fig.~\ref{fig:chempot}.
In the outer crust $\nu_n<0$, since all neutrons are bound to nuclei. The neutron drip is defined as the density at which the first neutron drips out of nuclei or equivalently by the condition that the neutron chemical potential $\nu_n > 0$. Fig.~\ref{fig:chempot} only shows this positive case.
The small dispersion of the curves in the lower left side of Fig.~\ref{fig:chempot} reflects the uncertainties in the position the outer-inner crust transition. As we already commented in the discussion of Fig.~\ref{fig:EOS}, the position of the inner-outer crust transition is better defined with Skyrme models than with $\ChiEFT$ ones. This indicates that the inner-outer crust transition is mainly determined by the experimental nuclear masses. As the density increases above the drip point, the neutron fluid contribution becomes more important. As a consequence a convergence of the $\ChiEFT$ models is observed, while the uncertainties on the Skyrme models predictions are getting larger and larger. As seen in the previous figures, we note that F0 and SLy5 (LNS5 and SGII) predicts higher (lower) values than the other models, while RATP, BSK14 and BSK16 predicts $\nu_n$ compatible with the $\ChiEFT$ models.

\begin{figure}[t]
\centering
\includegraphics[scale=0.31]{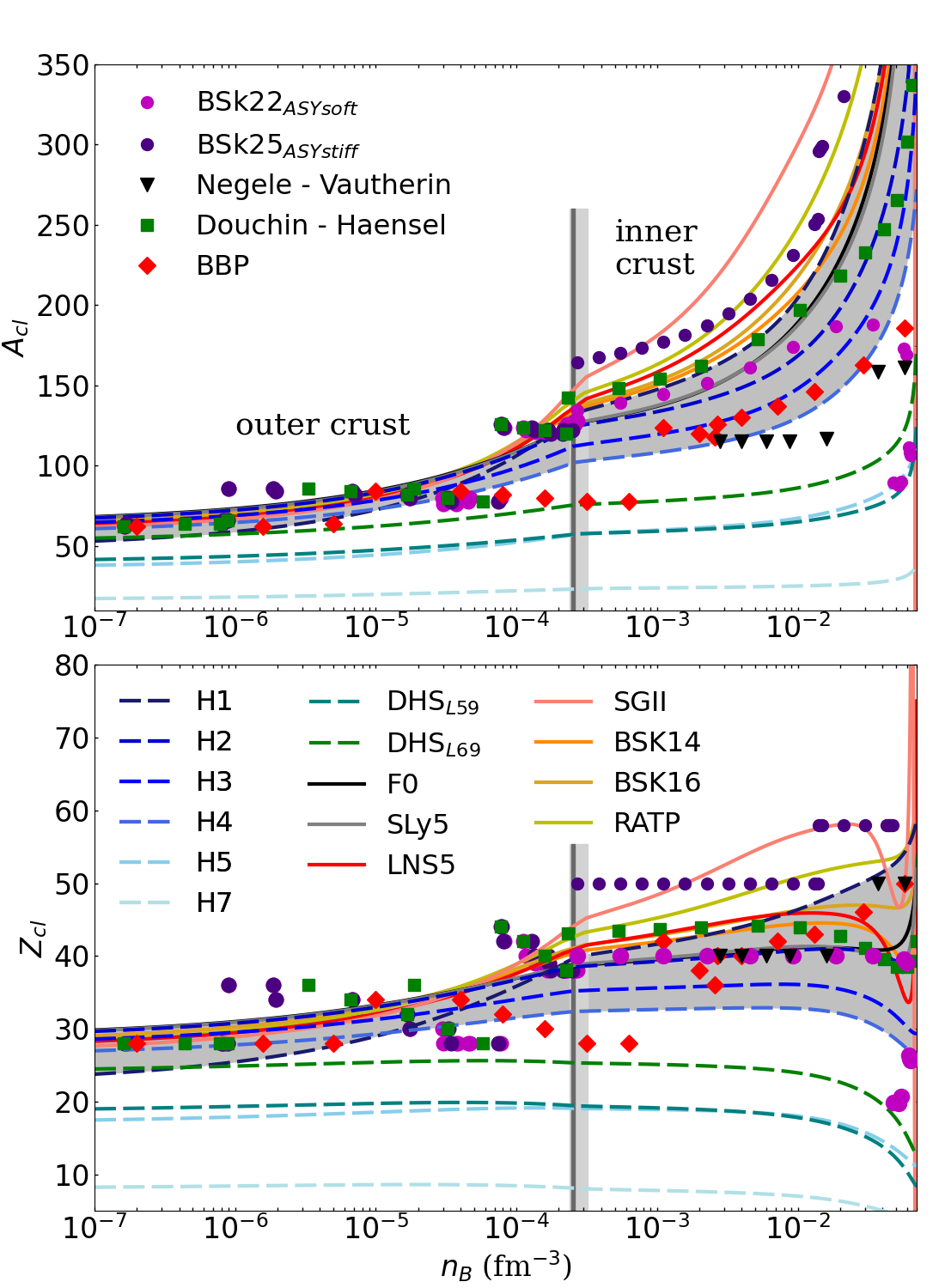}
\caption{Cluster composition, $A_\cl$ (top) and $Z\cl$ (bottom), for the eight $\ChiEFT$ Hamiltonians and the seven Skyrme models. The silver band shows our best results including H1-H4.}
\label{fig:composition}
\end{figure}

The cluster composition in the inner and outer crust is shown in Fig.~\ref{fig:composition}. Our best predictions from the $\ChiEFT$ models H1-H4 are bounded by the gray band. The Asy-Stiff models such as H5, H7, DHS$_{L59}$ and DHS$_{L69}$ predict lower $Z_\cl$ than the gray band, and the Asy-Soft models such as RATP, LNS5, SGII, BSK16 predict larger $Z_\cl$ than the gray band. Other models are compatible with the gray band. Note however that H5 and H7 predict much lower $A$ and $Z$ than the other $\ChiEFT$ ones. This is most probably due to the fact that these two hamiltonians predict the lowest values for $n_\sat$ compared to other $\ChiEFT$ predictions, see Tab.~\ref{tab:empirical:hamiltonians}. Since the cluster radius is mostly controlled by the virial condition, a lower value for $n_\sat$ induces a lower value for $A$. In addition, the cluster asymmetry parameter $I_\cl$ is predicted to be quite model independent, see our study in Ref.~\cite{Grams:2021b} for instance, a lower value for $A$ induces a lower value for $Z$.

An interesting feature appears close to the core-crust transition density: the behavior of $A_\cl$ and $Z_\cl$ can be quite different from one model to another. This is reflecting the important role played by the isospin asymmetry parameter $p_\surf$, which is fixed to be $p_\surf=3$ in our present study. The role of $p_\surf$ on the core-crust transition density has been discussed in Refs.~\cite{Carreau2019a,Grams:2021b}.
It is indeed a parameter which is difficult to determine, since finite nuclei do not explore isospin asymmetries large enough to be impacted by the value of $p_\surf$. This parameter could be determined from a slab configuration calculation exploring large asymmetries. In the present study, we however prefer to fix $p_\surf$ for simplicity, but a more accurate estimation of the core-crust transition density shall include a proper adjustment of the parameter $p_\surf$.

\begin{figure}[t]
\centering
\includegraphics[scale=0.35]{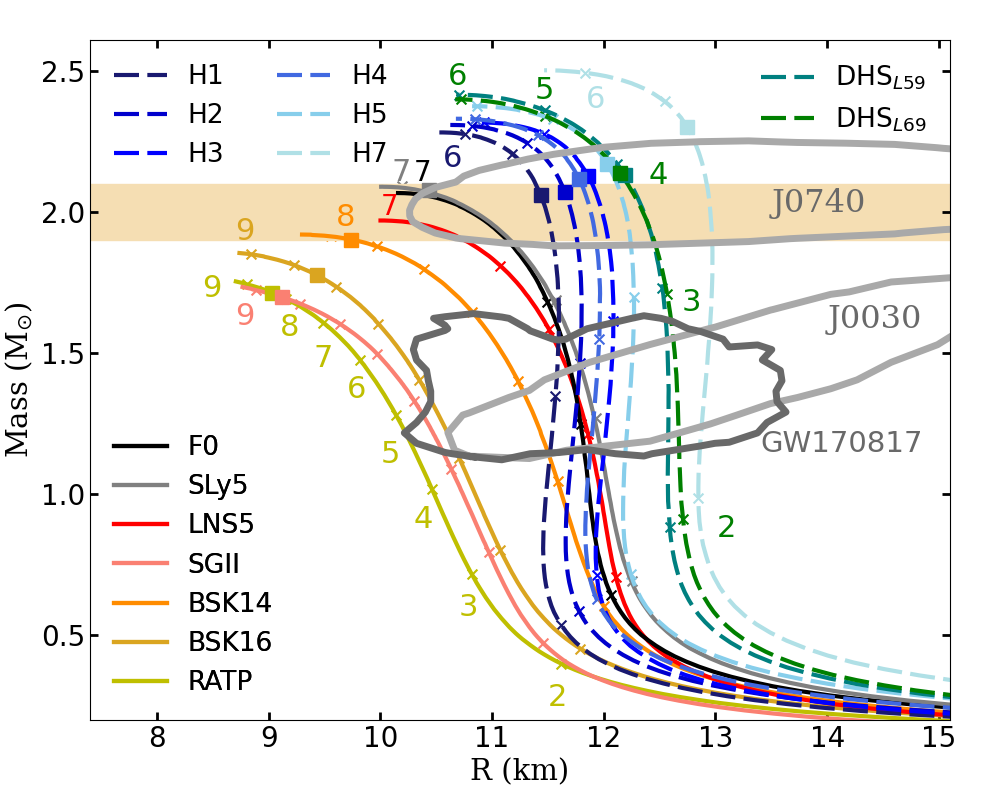}
\caption{Neutron star MR diagram, for the eight $\ChiEFT$ Hamiltonians and the seven Skyrme models. The squares indicate the density at which causality is violated. The light crosses mark the central density in units of $n_\sat$: the numbers given in the diagram indicate the value of the central density in units of $n_\sat$. Gray contours show the NICER observations for the pulsars J0030 (obtained from Ref.~\cite{Riley19}) and J0740 (obtained as an average of the analyses of Refs.~\cite{Miller21} and~\cite{NICER2021}) together with LIGO/Virgo observation GW170817~\cite{LIGOScientific:2018cki}.}
\label{fig:mr}
\end{figure}

We also compare our results with other predictions in Fig.~\ref{fig:composition}, such as the Bruxelles-Montreal Hartree-Fock Bogoliubov calculation (BSk22 and BSk25)~\cite{Pearson18}, the Negele-Vautherin Hartree-Fock calculation~\cite{Negele1973}, the Douchin-Haensel CLDM calculation~\cite{DouchinHaensel2000,DouchinHaensel2001,HaenselPichon94}, and finally the original BBP model~\cite{bbp1971}.
The oldest calculations, BBP, Negele-Vautherin and Douchin-Haensel, do not always overlap with our gray band, since they were performed before the recent $\ChiEFT$ achievement. The BSk22 model overlaps pretty well with our gray band, reflecting the good agreement already noticed for the symmetry energy. The model BSk25, which is stiffer than the $\ChiEFT$ H1-H4 models predicts larger $Z_\cl$ than us. Note also that while shell effects are absent from our calculation, our best predictions (gray band) is compatible with models which have them, e.g. BSk22 or Negele-Vautherin. This shows that while shell effects are important to get accurate $Z_\cl$, the actual value for $Z_\cl$ is still largely influenced by the contribution from the bulk term in the CLDM, in particular by the symmetry energy at low-density. The leptodermous expansion, which has been shown to provide a good ordering of the contribution of the different terms in the mass formula in NS crust, suggests that shell effects are comparable with curvature terms. In other words, shell effects are certainly important for accurate predictions of the crust properties, but they are not the main ingredient to understand the origin of the main uncertainties in these properties. The main source of uncertainties are originating from the bulk term, and in more detail by the low-density NM properties.

Finally, the mass-radius relations for all the considered models are given in Fig.~\ref{fig:mr}. The crosses on the curves mark the central density in units of $n_\sat$. Note that the central density at the maximum (causal) mass for the Skyrme models is of the order 7-8$n_\sat$.
Most of the Skyrme models considered here are softer and explore higher central densities than the $\ChiEFT$ ones (5-6$n_\sat$). 
Note that the MR relations require extrapolations above the $\ChiEFT$ breakdown density. We indeed recall that we have fixed the values of the parameters $Q$ and $Z$ in such a way that these models reach $2M_\odot$. Therefore the high density behavior of the $\ChiEFT$ EoS is mostly controlled by this prescription and is only mildly impacted by the $\ChiEFT$ properties at low density.
It is interesting to remark that all our EoSs satisfy the GW170817 constraint obtained from Ref.~\cite{LIGOScientific:2018cki} and shown as a dark-gray contour (90 $\%$ confidence level). The two light-gray contours in Fig.~\ref{fig:mr} represent the NICER measurements for the pulsars J0030~\cite{Riley19} and J0740~\cite{Miller21,NICER2021}. 
We see that all $\ChiEFT$ models together with the Skyme LNS5, F0 and SLy5 respect GW and NICER constraints, while BSK14 respects GW and only J0030. The other Skyrme models BSK16, SGII and RATP are outside the NICER contours and respect, but very marginally, the GW contour.
In addition, these three Skyrme models are very soft and fail to reproduce the observed lower limit of M$_\tov$. They are however stiff enough to predict M$_\tov\ge 1.7$~M$_\odot$, above the canonical mass NS. Despite the fact that they do not reach the observational constraint at around $2M_\odot$, we keep these EoSs since the densities at which they fail are well above the densities where nucleonic models can be trusted (3-4$n_\sat$). These EoSs certainly undergo a phase transition in their densest regions that we leave open for future studies.

\begin{figure}[t]
\centering
\includegraphics[scale=0.35]{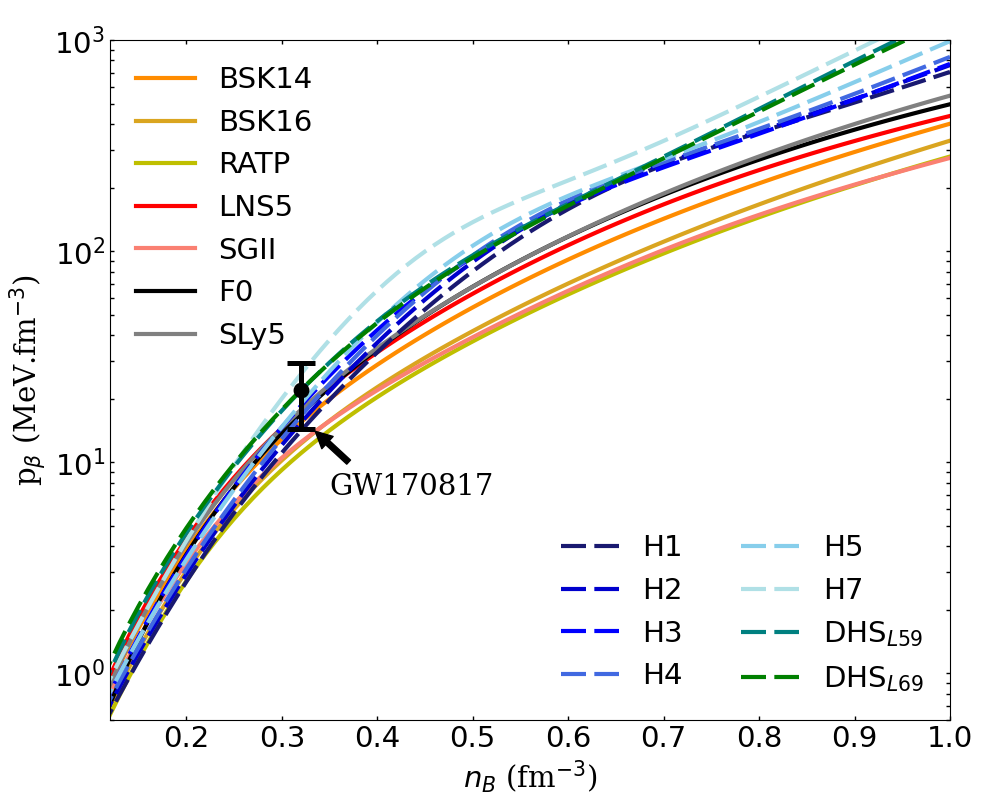}
\caption{Pressure versus baryon density (in fm$^{-3}$) at $\beta$-equilibrium, for the eight $\ChiEFT$ Hamiltonians and the seven Skyrme models. The error bar correspond to the inference for the pressure at twice saturation density from the LIGO/Virgo observation GW170817~\cite{LIGOScientific:2018cki}.}
\label{fig:eoslog}
\end{figure}

\begin{table}[t]
\centering
\caption{Neutron drip density $n_{\rm drip}$, crust-core transition density $n_{\rm cc}$, density at which the symmetry energy becomes negative $n_\sym$ and causal density $n_{\rm causal}$ above which causality is broken. For some Skyrme models (F0, LNS5 and SLy5) the maximum mass is reached before causality is broken. For some models (BSK16, LNS5, RATP and SGII) the softening of the EoS is explained from the low value of $n_{\rm sym}$.}
\label{tab:ndripNccNmaxCausal}
\tabcolsep=0.09cm
\def\arraystretch{1.6}
\begin{tabular}{lccccc}
\hline\noalign{\smallskip}
Model & $n_{\rm drip}$ & $n_{\rm cc}$ &  $n_{\rm sym}$ & $n_{\rm causal}$ \\
 &($\times 10^{-4}$ fm$^{-3}$) & ($\times 10^{-2}$ fm$^{-3}$) &(fm$^{-3}$) &(fm$^{-3}$) \\
\hline\noalign{\smallskip}
H1$_\mm$&  2.60 & 9.60   & - & 0.68  \\
H2$_\mm$ &  2.94 & 8.67  & - & 0.65 \\
H3$_\mm$ &  2.44 & 8.45  & - & 0.65 \\
H4$_\mm$&  2.52 & 9.06  & - & 0.65  \\
H5$_\mm$&  2.85 & 8.70  & - & 0.64  \\
H7$_\mm$ &  3.22 & 8.31  & - & 0.57 \\
DHS$^{L59}_\mm$ &  2.60  & 7.21 & -  & 0.62  \\
DHS$^{L69}_\mm$ &  2.37 & 7.32 & -  & 0.64 \\
\noalign{\smallskip}\hline\noalign{\smallskip}
BSK14$_\mm$  & 2.52 & 7.62  & - & 1.21 \\
BSK16$_\mm$  & 2.52 & 8.46  & 0.68 & 1.20 \\
F0$_\mm$  & 2.52 & 8.33  & - & - \\
LNS5$_\mm$  & 2.60 & 7.04  & 1.18 & - \\
RATP$_\mm$  & 2.52 & 8.60 & 0.63  & 1.34 \\
SGII$_\mm$  & 2.60 & 6.53 & 0.78  & 1.35 \\
SLY5$_\mm$  & 2.44 & 7.63 & -  & 1.03 \\
\noalign{\smallskip}\hline
\end{tabular}
\end{table}

The Skyrme models explored here are all quite soft at high density, see Fig.~\ref{fig:eoslog}. The soft Skyrme interactions which are marginal for the GW170817 contour shown in Fig.~\ref{fig:mr} (BSK16, SGII and RATP) are also marginal for the prediction of the pressure at $2n_\sat$ extracted from the analysis of GW based on agnostic EoS~\cite{LIGOScientific:2018cki}. It shall however be noticed that these interactions overlap with the GW contour for densities in the range $4$ to $5n_\sat$, so quite larger than the constraint inferred at $2n_\sat$. Even the stiffer models shown in Fig.~\ref{fig:mr} pass through the GW contour for densities above $2n_\sat$. The GW inference for the pressure at $2n_\sat$ shall therefore be considered cautiously.

The softening of BSK16, SGII and RATP EOSs is related to the bending down of the symmetry energy, see inset in Fig.~\ref{fig:enerSym}.
Note that in Fig.~\ref{fig:mr} the EOS for these three cases are replaced by the NM EOS for densities $n>n_\sym$, where $n_\sym$ is the density at which $e_\sym=0$, see Tab.~\ref{tab:ndripNccNmaxCausal}.
The other Skyrme EOSs, BSK14, LNS5, SLy5 and F0, are softer than the $\ChiEFT$ ones for densities $n_B\gtrsim 0.3$-$0.4$~fm$^{-3}$ as shown in Fig.~\ref{fig:eoslog}. 
It is then interesting to study the consequence of such soft EoSs on the properties of neutron stars.
The first two columns of Tab. ~\ref{tab:ndripNccNmaxCausal} summarize the boundaries for each EoS. The first column shows the density transition between the outer and inner crust $n_{\rm drip}$ while the second shows the crust-core transition $n_\cc$. In the two last columns we show the limit of validity for the EoS that either present negative values for the symmetry energy at high densities ($n_\sym$) or break causality ($n_{\rm causal}$).

Finally, we observe in Fig.~\ref{fig:eoslog} a change of slope of the pressure for densities of about 0.4 to 0.6~fm$^{-3}$, similar to the one observed in Ref.~\cite{Annala:2019}. The authors of Ref.~\cite{Annala:2019} interpreted this softening as a sign of the presence of quark matter in massive neutron stars. This cannot be our case since we describe purely nucleonic matter. This apparent contradiction is discussed in Ref.~\cite{Somasundaram2021b}, where it is shown that the bending down of the pressure at these densities is not necessarily correlated with the reduction of the sound speed, as it should be for a phase transition to quark matter. In the present approach, we indeed illustrate that the change of slope of the pressure does not necessarily reflect the onset of a phase transition to quark matter. The bending of the pressure is compatible with matter composed of nucleons only, where the many-body correlations at high density contribute to soften the EoS.

\section{Conclusions}

In the present analysis, we have explored the predictions of fifteen models, including seven Skyrme and eight $\ChiEFT$ Hamiltonians. We have focused our study on the properties of the crust, for which we have employed a CLDM approach. This approach is adequate for the understanding of the origin of the model dependence in the prediction for the NS crust properties. We have indeed found that there are two important features which govern, at first order, our predictions: i) the ability of the models to reproduce the nuclear experimental masses over the nuclear table, and ii) the low-density energy per particle in NM. For the models which satisfy feature (i), the condition on the low-density energy per particle in NM is largely equivalent to the one on the low-density properties of the symmetry energy.

Crust properties have been analyzed in terms of bulk and FS contributions, where the bulk term explore the uncertainties in the uniform matter prediction, e.g. the low-density NM, and the FS terms are optimized to reproduce experimental nuclear masses. The two main sources of uncertainties for the NS crust properties are then well identified. We have observed that some quantities are more sensitive to the feature (i), e.g. cluster composition $A_\cl$ and $Z_\cl$, while others are more impacted by the feature (ii), e.g. energy per particle, pressure, sound speed, volume fraction, electron fraction $Y_e$, neutron chemical potential, mentioning only the observables that we have analyzed. In a previous analysis, we have also illustrated the role of FS terms, in particular the effect of the parameter $p_\surf$ controlling the isospin dependence of the surface energy for large values of the isospin parameter.

For all the fifteen unified EoS presented here, we have generated tables in the CompOSE format.  Many nucleonic models presented here predict  M$_\tov$ above the limit set by radio-observations of about 2M$_\odot$, except a few Skyrme EOSs. The central densities explored by these soft EoS are however so large that the nucleon model presented here shall be replaced by a model with exotic degrees of freedom. The description of phase transitions is out of the scope of the present study, but it could be performed on top of the EOS we provide. Models with phase transition will be studied by the authors in future works.

In summary, our study illustrates the importance of constructing unified models for NS EoS. Further extensions are currently considered, investigating contributions from smaller terms in the leptodermous expansion, such as for instance the pairing term or the shell effects.

\begin{acknowledgements}
We thank N. Chamel for providing the nuclear
configurations below saturation predicted by BSk22 to BSK25.
G.G., J.M. and R.S. are supported by CNRS grant PICS-08294 VIPER (Nuclear Physics for Violent Phenomena in the Universe), the CNRS IEA-303083 BEOS project, the CNRS/IN2P3 NewMAC project, and benefit from PHAROS COST Action MP16214. S.R. is supported by Grant No. DE-FG02-00ER41132 from the Department of Energy , and the Grant No. PHY-1430152 (JINA Center for the Evolution of the Elements), and PHY-1630782 (Network for Neutrinos, Nuclear Astrophysics, and Symmetries (N3AS)) from the National Science Foundation. This work is supported by the LABEX Lyon Institute of Origins (ANR-10-LABX-0066) of the \textsl{Universit\'e de Lyon} for its financial support within the program \textsl{Investissements d'Avenir} (ANR-11-IDEX-0007) of the French government operated by the National Research Agency (ANR).
\end{acknowledgements}


\bibliographystyle{spphys}       
\bibliography{ref}   

%
%

\end{document}